\documentclass[english,prl, notitlepage, onecolumn, longbibliography]{revtex4-1}
\usepackage[T1]{fontenc}
\usepackage[latin9]{inputenc}
\usepackage[active]{srcltx}
\usepackage{color}
\usepackage{array}
\usepackage{bm}
\usepackage{amsmath}
\usepackage{graphicx}

\makeatletter
\providecommand{\tabularnewline}{\\}
\usepackage{amsmath}
\usepackage{subfigure}
\usepackage{graphicx, mathtools}
\usepackage[normalem]{ulem}
\usepackage[colorlinks=true,linkcolor=MidnightBlue,urlcolor=MidnightBlue,citecolor=MidnightBlue,anchorcolor=MidnightBlue]{hyperref}\usepackage[dvipsnames]{xcolor}
\definecolor{myGreen}{rgb}{0.0, 1, 10.1}
\definecolor{myBlue}{rgb}{0.9, 0.2, 0}

\makeatother

\usepackage{babel}
\begin{document}

\title{Fluctuating hydrodynamics and the Brownian motion of an active colloid
near a wall }

\author{Rajesh Singh}
\email{rsingh@imsc.res.in}

\affiliation{The Institute of Mathematical Sciences-HBNI, CIT Campus, Chennai
600113, India}

\author{R. Adhikari}
\email{rjoy@imsc.res.in}

\affiliation{The Institute of Mathematical Sciences-HBNI, CIT Campus, Chennai
600113, India}
\begin{abstract}
The traction on the surface of a spherical active colloid in a thermally
fluctuating Stokesian fluid contains passive, active, and Brownian
contributions. Here we derive these three parts systematically, by
``projecting out'' the fluid using the boundary-domain integral
representation of slow viscous flow. We find an exact relation between
the statistics of the Brownian traction and the thermal forces in
the fluid and derive, thereby, fluctuation-dissipation relations for
every irreducible tensorial harmonic traction mode. The first two
modes give the Brownian force and torque, from which we construct
the Langevin and Smoluchowski equations for the position and orientation
of the colloid. We emphasize the activity-induced breakdown of detailed
balance and provide a prescription for computing the configuration-dependent
variances of the Brownian force and torque. We apply these general
results to an active colloid near a plane wall, the simplest geometry
with configuration-dependent variances, and show that the stationary
distribution is non-Gibbsian. We derive a regularization of the translational
and rotational friction tensors, necessary for Brownian dynamics simulations,
that ensures positive variances at all distances from the wall. The
many-body generalization of these results is indicated.\\\\DOI: \href{http://www.tandfonline.com/doi/abs/10.1080/17797179.2017.1294829?journalCode=tecm20/a}{10.1080/17797179.2017.1294829}
\end{abstract}
\maketitle

\section{Introduction}

There has been a renewal of interest in the study of colloids with
``active'' boundaries, on which the usual no-slip boundary condition
is replaced by one involving a ``slip velocity''. This slip is the
macroscopic manifestation of microscopic non-equilibrium processes
in a thin boundary layer surrounding the colloid. Classic examples
of slip-driven motion are the multitude of phoretic phenomena including
electro-, thermo- and diffusio-phoresis and the motion of ciliated
microorganisms \cite{anderson1989colloid,ebbens2010pursuit}. More
recently, the slip model has been adapted to provide an effective
description of flagellated microorganisms \cite{ghose2014irreducible}.
It provides a very general framework for the dynamics of phenomena
where colloidal motion occurs without external influence.

The traction, that is the force per unit area, on the surface of an
active colloid has three components: the Stokes drag proportional
to the rigid body motion, the active thrust proportional to the slip
velocity and the Brownian stress proportional to the temperature and
the viscosity of the fluid. When the inertia of the colloid is negligible,
its rigid body motion is obtained by setting the net force and net
torque due to these tractions to zero. A central problem, then, is
to derive explicit expressions for each part of the traction and to
obtain, thereby, the force, the torque and its remaining moments. 

At low Reynolds number and at finite temperature the Cauchy stress
in the fluid obeys the fluctuating Stokes equation, a linear stochastic
partial differential equation containing zero-mean Gaussian random
fluxes with variances determined by the fluctuation-dissipation relation.
These represent thermal fluctuations in the fluid. The solenoidal
fluid velocity obeys the slip boundary condition at the colloid-fluid
boundary. For a given rigid body motion and slip, the solution of
the fluctuating Stokes equation provides the Cauchy stress in the
fluid and, hence, the traction on the boundary. This solution for
the traction only contains the boundary condition and the random fluxes;
the fluid is, therefore, ``projected out''. The Brownian forces
and torques on the colloid are the first and second antisymmetric
moments of the stochastic part of the traction. Their variance, by
linearity of the governing equations, is proportional to the variance
of the random fluxes. The Langevin equations for the position and
orientation of the colloid follow straightforwardly by inserting the
expressions for the net force and torque in the corresponding Newton's
equation. 

Fox and Uhlenbeck were the first to derive the Langevin equation for
the position of a \emph{passive} spherical colloid from the fluctuating
hydrodynamic equations for the fluid \cite{fox1970contributions}.
The fluid was taken to satisfy the no-slip boundary condition on the
surface of the colloid and to be quiescent at the remote boundaries.
The Lorentz reciprocal identity was used to relate the deterministic
(Stokes) and stochastic (Einstein) parts of the force and to derive,
thereby, the fluctuation-dissipation relation for the Brownian  force
from that of the random fluxes in the fluid. This approach was extended
by several authors to include fluid inertia, particle inertia, Brownian
fluxes at the colloid-fluid boundary and to many colloidal particles
\cite{hauge1973fluctuating,bedeaux1974brownian,beenakker1983b,noetinger1990fluctuating,roux1992brownian}.
Zwanzig, in an earlier contribution, derived the variance of the Brownian
force on a spherical colloid in an unbounded fluid using the Faxén
relation. The use of this variance in the Green-Kubo relation for
the transport coefficient recovered the correct value of the Stokes
friction \cite{zwanzig1964hydrodynamic}.

In this contribution, we derive the traction on the surface of an
$active$\emph{ }colloidal particle near a \emph{plane wall} by projecting
out the fluid degrees of freedom. Our derivation differs in three
important ways from previous work. First, we derive the complete distribution
of the Brownian traction, and not just its first two moments as has
been customary. This provides, in addition to the Brownian force and
torque, the Brownian stresslet, a quantity important in suspension
rheology. Second, the no-slip boundary condition at the colloid-fluid
boundary is \emph{replaced} by the slip boundary condition. This introduces
an additional uncompensated source of dissipation and is the source,
as we shall see, of the breakdown of detailed balance in the Langevin
equations. Third, a no-slip boundary condition is \emph{introduced}
at the plane wall. This results in Stokes drags and active thrusts
that depend on the distance of the colloid from the wall. The configuration-dependent
friction requires, by the fluctuation-dissipation relation, configuration-dependent
Brownian forces and torques and necessitates a ``prescription''
to render the Langevin equations mathematically meaningful. The solution
to this old chestnut, the so-called Itô-Stratonovich dilemma, has
been provided by multiple authors on multiple occasions but tends
to be forgotten \cite{lau2007state,volpe2010influence,volpe2016effective,mannella2011comment}.
The procedure is to adiabatically eliminate the momentum, considered
as a fast variable, from underdamped Langevin equations (where no
such dilemma exists). The interpretation of the Brownian forces and
torques in the overdamped Langevin equations is then unambiguous,
though not necessarily conforming to either the Itô or Stratonovich
prescriptions \cite{van1981ito,gardiner1985handbook,van1992stochastic,klimontovich1990ito,klimontovich1994nonlinear}.
Due to the linearity of the governing equations, this method of imputing
meaning to configuration-dependent Brownian forces and torques remains
valid in the presence of activity, as we show below. To the best of
our knowledge, ours is the first systematic derivation of the Langevin
equations for an active (and, as a special case, of a passive) colloid
when the Stokes friction is configuration-dependent.

The remainder of the paper is organized as follows. In section \ref{sec:integral-equation},
we transform the fluctuating Stokes equation to its boundary-domain
integral representation and hence obtain a linear integral equation
for the traction on the colloid-fluid boundary. A formal solution,
expressed in terms of the inverse of the single-layer operator of
the integral equation (see below), clearly identifies the passive,
active and Brownian parts of the traction. In section \ref{sec:Force-and-torque},
we consider a spherical colloid and provide an explicit solution to
the boundary-domain integral equation. We derive, through a Galerkin
discretization, an equivalent linear \emph{algebraic} system for the
coefficients of the expansion of the traction in a complete orthogonal
basis of tensorial spherical harmonics. The solution of the linear
system shows that the each tensorial coefficient of the Brownian traction
is a zero-mean Gaussian random variable and provides their variances
in terms of the variance of the stochastic term in the fluctuating
Stokes equation. Variances of the Brownian force, torque and stresslet
follow immediately. In section \ref{sec:Langevin-equations- }, we
use the previous results to construct the overdamped Langevin equations
for the position and orientation of the colloid. The Smoluchowski
equations, corresponding to the prescription for the Brownian forces
and torques provided by the adiabatic elimination procedure, are then
presented and the activity-induced breakdown of the fluctuation-dissipation
relation is pointed out. In section \ref{sec:wall-Bounded}, we specialize
to the case of Stokes flow bounded by a plane wall and provide a leading
order solution in terms of the Green's function first identified by
Lorentz and subsequently derived systematically by Blake \cite{blake1971c}.
We show that the usual barometric distribution of height is no longer
the stationary solution, a consequence of the breakdown of detailed
balance. We also provide a regularization of the friction tensors,
for what would correspond to heights in which the sphere overlaps
with the wall. This ensures positive variances and is a necessary
ingredient in Brownian dynamics simulations of spheres without hard
steric potentials. Finally, the extension of the above results to
many active colloids is indicated.

\section{Boundary-domain integral representation of fluctuating Stokes flow
\label{sec:integral-equation}}

We consider, in this section, the motion of an active colloid of arbitrary
shape in an incompressible fluid of viscosity $\eta$ at a temperature
$k_{B}T$. The boundary condition induces a local force per unit area
$\boldsymbol{f}=\mathbf{\hat{\boldsymbol{\rho}}}\cdot\boldsymbol{\sigma}$
on the colloid boundary, where $\boldsymbol{\sigma}$ is the Cauchy
stress in the fluid and \textbf{$\mathbf{\boldsymbol{\hat{\rho}}}$}
is the unit surface normal \cite{landau1959fluid}. We shall henceforth
refer to $\boldsymbol{f}$ as the traction. In addition, the colloid
may also be acted upon by body forces $\mathbf{F}^{P}$ and body torques
$\mathbf{T}^{P}$. In the absence of inertia of both particle and
fluid, Newton's equations for the colloid reduces to instantaneous
balance of the surface forces (and torques) with the body forces (and
torques):
\begin{alignat}{1}
\int\boldsymbol{f}\,dS+\mathbf{F}^{P}=0,\qquad & \int\mathbf{\boldsymbol{\rho}}\times\boldsymbol{f}\,dS+\mathbf{T}^{P}=0.\label{eq:newtons}
\end{alignat}
These are overdamped Langevin equations when $\boldsymbol{f}$ contains
Brownian contributions. 

To obtain the traction $\boldsymbol{f}$ it is necessary to know the
flow field $\mathbf{u}$. At low-Reynolds number this satisfies the
Stokes equation \cite{happel1965low}\begin{subequations}\label{eq:stokesPart}
\begin{alignat}{1}
\boldsymbol{\nabla}\cdot\mathbf{u}=0,\quad\boldsymbol{\nabla}\cdot\boldsymbol{\sigma}+\boldsymbol{\xi}=0,\qquad & \text{in }V,\\
\mathbf{u}=\text{\ensuremath{\boldsymbol{v}}},\qquad & \text{on }S,
\end{alignat}
\end{subequations}where $\bm{\sigma}=-p\boldsymbol{I}+\eta(\bm{\nabla}\mathbf{u}+(\bm{\nabla}\mathbf{u})^{T})$
is the Cauchy stress, $p$ is the fluid pressure, $\boldsymbol{\xi}$
is the thermal force acting on the fluid, $\boldsymbol{v}$ is the
boundary velocity and $S$ is the surface of the colloid and $V$
is the domain of flow. The thermal force is a zero-mean Gaussian random
field whose variance is given by the fluctuation-dissipation relation
\begin{equation}
\left<\int\mathbf{u}^{\mathcal{D}}(\mathbf{r})\cdot\boldsymbol{\xi}(\mathbf{r},\,t)\,dV\int\mathbf{u}^{\mathcal{D}}(\mathbf{r}')\cdot\boldsymbol{\xi}(\mathbf{r}',\,t')\,dV'\right>=2k_{B}T\,\dot{\mathcal{E}}(\mathbf{\mathbf{u}^{\mathcal{D}})}\delta(t-t'),\label{eq:FDT-Hauge}
\end{equation}
where $\mathbf{u}^{\mathcal{D}}$ is any flow field that satisfies
the rigid body boundary condition on $S$ and $\dot{\mathcal{E}}$,
the rate of dissipation of the fluid kinetic energy due to rigid body
motion, is given by
\begin{equation}
\dot{\mathcal{E}}(\mathbf{\mathbf{u}^{\mathcal{D}})}=\eta\int\left[\boldsymbol{\nabla}\mathbf{u}^{\mathcal{D}}+\left(\boldsymbol{\nabla}\mathbf{u}^{\mathcal{D}}\right){}^{T}\right]{}^{2}dV.\label{eq:dissipation-W}
\end{equation}
This manner of describing the thermally fluctuating fluid, due to
Hauge and Martin-Löf \cite{hauge1973fluctuating}, is specially suited
for flows with boundaries. The addition of a random flux, the more
conventional manner of description first introduced by Landau and
Lifshitz \cite{landau1959fluid}, contains ambiguities in the presence
of boundaries and is best used, therefore, when the fluid is unbounded
in all directions \cite{hauge1973fluctuating}. 

The key property of the above problem, which makes it possible to
eliminate the fluid degrees of freedom \emph{exactly}, is linearity.
It is most clearly expressed in terms of the boundary-domain integral
representation of slow viscous flow which provides the fluid flow
in the bulk in terms of the boundary condition and the thermal force.
The traction $\boldsymbol{f}$ is obtained as a solution of the boundary-domain
integral equation
\begin{alignat}{1}
\frac{1}{2}v_{\alpha}(\mathbf{r})= & \int G_{\alpha\beta}(\mathbf{r},\,\mathbf{r}')\,\xi_{\beta}(\mathbf{r}')\,dV'-\int G_{\alpha\beta}(\mathbf{r},\,\mathbf{r}')f_{\beta}(\mathbf{r}')\,d\mathrm{S'}+\int K_{\beta\alpha\gamma}(\mathbf{r},\,\mathbf{r}')\hat{\rho}'_{\gamma}v_{\beta}(\mathbf{r}')\,d\mathrm{S}',\label{eq:BIE-full}
\end{alignat}
where $G_{\alpha\beta}(\mathbf{r},\,\mathbf{r}')$ is a Green's function
of the Stokes equation and $K_{\alpha\beta\gamma}(\mathbf{r},\,\mathbf{r}')$
is the associated stress tensor. These kernels satisfy the Stokes
system \begin{subequations}
\begin{alignat}{1}
\nabla_{\alpha}G_{\alpha\beta}(\mathbf{r},\,\mathbf{r'})=0,\qquad-\nabla_{\alpha}P_{\beta}(\mathbf{r},\,\mathbf{r'})+\eta\nabla^{2}G_{\alpha\beta}(\mathbf{r},\,\mathbf{r'})=-\delta\left(\mathbf{r}-\mathbf{r'}\right)\delta_{\alpha\beta},\\
K_{\alpha\beta\gamma}(\mathbf{r},\mathbf{r'})=-\delta_{\alpha\gamma}P_{\beta}(\mathbf{r},\,\mathbf{r'})+\eta\left(\nabla_{\gamma}G_{\alpha\beta}(\mathbf{r},\,\mathbf{r'})+\nabla_{\alpha}G_{\beta\gamma}(\mathbf{r},\,\mathbf{r'})\right),\quad\,\,
\end{alignat}
\end{subequations}where $P_{\alpha}$($\mathbf{r},\,\mathbf{r}')$
is the pressure vector. Implicit in the above is a choice of Green's
function that satisfies no-slip boundary conditions at any boundary
of the fluid that is not part of $S$. 

Defining the single-layer and double-layer integral operators $\boldsymbol{G}$
and $\boldsymbol{K}$, which act, respectively, on tractions and velocities,
as
\begin{equation}
\boldsymbol{G}\cdot\boldsymbol{f}=\int\mathbf{G}(\mathbf{r},\,\mathbf{r}')\cdot\boldsymbol{f}(\mathbf{r}')\,d\mathrm{S',}\qquad\boldsymbol{K}\cdot\boldsymbol{v}=\int\mathbf{\hat{\bm{\rho}}'\cdot K}(\mathbf{r},\,\mathbf{r}')\cdot\boldsymbol{v}(\mathbf{r}')\,d\mathrm{S}',
\end{equation}
and a Brownian velocity field $\boldsymbol{w}$ as
\begin{equation}
\boldsymbol{w}=\int\mathbf{G}(\mathbf{r},\,\mathbf{r}')\cdot\boldsymbol{\xi}(\mathbf{r}')\,dV',\label{eq:Brownian-velocity-field}
\end{equation}
the solution of the boundary-domain integral equation can be expressed
formally in terms of the inverse of the single-layer integral operator
as
\begin{alignat}{1}
\boldsymbol{f}= & \,\boldsymbol{G}^{-1}\cdot\boldsymbol{w}+\boldsymbol{G}^{-1}\cdot\left(-\tfrac{1}{2}\boldsymbol{I}+\boldsymbol{K}\right)\cdot\boldsymbol{v}.\label{eq:linear-system}
\end{alignat}
This formal solution shows that: (i) the traction is a sum of a Brownian
contribution $\boldsymbol{\hat{f}}=\,\boldsymbol{G}^{-1}\cdot\boldsymbol{w}$
from the fluctuations in the fluid and a deterministic contribution
from the boundary condition, containing both the rigid body motion
$\boldsymbol{f}^{\mathcal{D}}=-\boldsymbol{G}^{-1}\cdot\boldsymbol{v}^{\mathcal{D}}$
and the active slip $\boldsymbol{f}^{\mathcal{A}}=\boldsymbol{G}^{-1}\cdot\left(-\tfrac{1}{2}\boldsymbol{I}+\boldsymbol{K}\right)\cdot\boldsymbol{v}^{\mathcal{A}}$
(ii) the Brownian  traction is a zero-mean Gaussian random variable
whose variance is linearly related to the variance of the thermal
force $\boldsymbol{\xi}$ and (iii) the variance of the Brownian  traction
can be determined from the inverse of the single-layer operator and
the fluctuation-dissipation relation for the thermal force. In the
next section, we provide a solution for the boundary-domain integral
equation in a basis adapted for a spherical active colloid and, thereby,
derive the explicit form of the traction in terms of generalized friction
tensors.

\section{Traction on a spherical active colloid \label{sec:Force-and-torque}}

We now consider a spherical colloid of radius $b$ whose center is
at \textbf{$\mathbf{R}$} and whose orientation is specified by the
unit vector $\mathbf{p}$. A point on the boundary is $\mathbf{r}=\mathbf{R}+\boldsymbol{\rho}$,
where $\boldsymbol{\rho}$ is the radius vector. The boundary velocity,
\textbf{$\boldsymbol{v}=\text{\ensuremath{\boldsymbol{v}}}^{\mathcal{D}}+\boldsymbol{v}^{\mathcal{A}}$},
is the sum of its rigid body motion \textbf{$\text{\ensuremath{\boldsymbol{v}}}^{\mathcal{D}}=\mathbf{V}+\boldsymbol{\Omega}\times\boldsymbol{\rho}$},
specified by the velocity $\mathbf{V}$ and angular velocity $\mathbf{\Omega}$,
and the active slip velocity $\boldsymbol{v}^{\mathcal{A}}$. The
only restriction on the active slip is that its integral over the
surface of the sphere is zero. This ensures conservation of mass in
the fluid. 

We choose the tensorial spherical harmonics $\mathbf{Y}^{(l)}(\bm{\hat{\rho}})=(-1)^{l}\rho^{l+1}\bm{\nabla}^{(l)}\rho^{-1}$,
where $\bm{\nabla}^{(l)}=\bm{\nabla}_{\alpha_{1}}\dots\bm{\nabla}_{\alpha_{l}}$,
as complete orthogonal basis functions on the sphere. In this basis,
the active slip is expanded as\textit{
\begin{equation}
\boldsymbol{v}^{\mathcal{A}}=\sum_{l=1}^{\infty}\frac{1}{(l-1)!(2l-3)!!}\,\mathbf{V}^{(l)}\cdot\mathbf{Y}^{(l-1)}(\bm{\hat{\rho}}).
\end{equation}
}The coefficients $\mathbf{V}^{(l)}$ are reducible Cartesian tensors
of rank $l$, with three irreducible parts of ranks $l,$ $l-1,$
and $l-2$, corresponding to symmetric traceless, antisymmetric and
pure trace combinations of the reducible indices. These irreducible
parts are $\mathbf{V}^{(l\sigma)}=\mathbf{P}^{(l\sigma)}\cdot\mathbf{V}^{(l)}$,
where the index $\sigma=s,\,a\text{ and }t$, labels the symmetric
irreducible, antisymmetric and pure trace parts of the reducible tensors
\cite{brunn1976effect,schmitz1980force,ghose2014irreducible,singh2015many,singh2016traction,singh2016crystallization}.
The operator $\mathbf{P}^{(ls)}=\boldsymbol{\Delta}^{(l)}$ extracts
the symmetric irreducible part, $\mathbf{P}^{(la)}=\boldsymbol{\Delta}^{(l-1)}\boldsymbol{\boldsymbol{\varepsilon}}$
the antisymmetric part and $\mathbf{P}^{(lt)}=\boldsymbol{\delta}$
the trace of the operand. Here $\mathbf{\Delta}^{(l)}$ is a tensor
of rank $2l$, projecting any $l$-th order tensor to its symmetric
irreducible form \cite{hess2015tensors}, $\boldsymbol{\varepsilon}$
is the Levi-Civita tensor and $\boldsymbol{\mathbf{\delta}}$ is the
Kronecker delta. 

The traction can be similarly expanded in the tensorial harmonic basis
as\textit{
\begin{equation}
\boldsymbol{f}=\sum_{l=1}^{\infty}\frac{2l-1}{4\pi b^{2}}\,\mathbf{F}^{(l)}\cdot\mathbf{Y}^{(l-1)}(\bm{\hat{\rho}}).
\end{equation}
}and, as with the velocity coefficients, each traction coefficient
at order $l$ has three irreducible parts indexed by $\sigma$. The
net force and the torque are given by first two irreducible coefficients
\begin{equation}
\text{\ensuremath{\int\boldsymbol{f}}\,d}S=\mathbf{F}^{(1s)},\qquad\int\boldsymbol{\rho}\times\boldsymbol{f}\,dS=b\mathbf{F}^{(2a)}.
\end{equation}
Note that each $\mathbf{V}^{(l)}$ and $\mathbf{F}^{(l)}$ have the
dimension of velocity and force respectively. 

It is convenient to express the traction as a sum of rigid body, active
and Brownian contributions, 
\begin{equation}
\boldsymbol{f}=\boldsymbol{f}^{\mathcal{D}}+\boldsymbol{f}^{\mathcal{A}}+\boldsymbol{\hat{f}},
\end{equation}
with corresponding expansion coefficients $\mathbf{F}^{\mathcal{D}(l)}$,
$\mathbf{F}^{\mathcal{A}(l)}$ and $\hat{\mathbf{F}}^{(l)}$. By linearity,
the three parts of the traction satisfy\emph{ independent }boundary
integral equations. Recalling that rigid body motion is an eigenvector
of the double-layer integral operator, these are\begin{subequations}\label{eq:BIE-3parts}
\begin{eqnarray}
V_{\alpha}+\epsilon_{\alpha\beta\gamma}\Omega_{\beta}\rho_{\gamma} & =-\int G_{\alpha\beta}^{\text{}}(\mathbf{r},\,\mathbf{r}')f_{\beta}^{\mathcal{D}}(\mathbf{r}')\,d\mathrm{S'},\qquad\quad\,\, & \text{(rigid body)}\label{eq:bie-passive}\\
\frac{1}{2}v_{\alpha}^{\mathcal{A}}(\mathbf{r})=-\int G_{\alpha\beta}(\mathbf{r},\,\mathbf{r}')f_{\beta}^{\mathcal{A}}(\mathbf{r}')\,d\mathrm{S'} & +\int K_{\beta\alpha\gamma}^{\text{}}(\mathbf{r},\,\mathbf{r}')\hat{\rho}'_{\gamma}v_{\beta}^{\mathcal{A}}(\mathbf{r}')\,d\mathrm{S}',\qquad\qquad & \text{(active)}\label{eq:bie-active}\\
0=\int G_{\alpha\beta}(\mathbf{r},\mathbf{r}')\,\mathbf{\xi}_{\beta}(\mathbf{r}')\,dV' & -\,\,\int G_{\alpha\beta}^{\text{}}(\mathbf{r},\,\,\mathbf{r}')\,\,\hat{f}_{\beta}(\,\mathbf{r}')\,\,d\text{S}'.\qquad\qquad & \text{(Brownian)}\label{eq:bie-brownian}
\end{eqnarray}
\end{subequations}Addition of the above three equations recovers
the boundary integral equation, Eq. (\ref{eq:BIE-full}), for the
net traction. 

The first integral equation for the Stokes drag has been well-studied
in the literature on suspension mechanics. The second integral equation
for the active thrust has been studied recently in the context of
active colloids in an athermal fluid \cite{singh2015many,singh2016crystallization,singh2016traction}.
The third integral equation for the fluctuating traction is studied
here for the first time. Physically, the fluctuating traction corresponds
to the distribution of surface forces necessary to keep the sphere
stationary in the incident Brownian  velocity field $\boldsymbol{w}(\mathbf{r)}$.
From this, it is particularly clear that the Brownian  traction is
a zero-mean Gaussian random variable whose variance is related to
that of $\boldsymbol{\xi}$. We now present explicit solutions for
each of the integral equations using Galerkin's method of discretization.
Linear algebraic systems are derived by inserting the basis expansions
on each side of the integral equations, weighting the result by another
basis function and integrating over the boundary. We refer the reader
to \cite{singh2015many,singh2016crystallization,singh2016traction}
where the procedure is explained in detail. 

\subsection{Rigid body traction}

The linear algebraic system for the rigid body contribution to the
traction, with the summation convention for repeated indices, is
\begin{alignat}{1}
 & -\boldsymbol{G}^{(l,\,l')}(\mathbf{R})\cdot\mathbf{F}^{\mathcal{D}(l')}=\mathbf{V}^{\mathcal{D}(l)}.\label{eq:linear-system-passive}
\end{alignat}
Here the matrix elements of the single-layer operator $\boldsymbol{G}(\mathbf{R)}$
in the tensorial harmonic basis are 
\begin{equation}
\boldsymbol{G}^{(l,\,l')}(\mathbf{R})=\frac{2l-1}{4\pi b^{2}}\frac{2l'-1}{4\pi b^{2}}\int\mathbf{Y}^{(l-1)}(\hat{\bm{\rho}})\mathbf{G}^{\text{}}(\mathbf{R}+\bm{\rho},\,\mathbf{R}+\bm{\rho}')\mathbf{Y}^{(l'-1)}(\hat{\bm{\rho}}')\,d\mathrm{S}\,d\mathrm{S'},
\end{equation}
and $l$-th tensorial harmonic coefficients of the rigid body motion
and the traction are, respectively, $\mathbf{V}^{\mathcal{D}(l)}$
and $\mathbf{F}^{\mathcal{D}(l)}$. Clearly, the only non-zero coefficients
of $\mathbf{V}^{\mathcal{D}(l)}$ are $\mathbf{V}$ and $\boldsymbol{\Omega},$
corresponding to $l\sigma=1s$ and $l\sigma=2a$ respectively. The
solution of the linear system is
\begin{alignat}{1}
\mathbf{F}^{\mathcal{D}(l\sigma)}= & -\,\boldsymbol{\gamma}^{(l\sigma,\,1s)}\cdot\mathbf{V}-\boldsymbol{\gamma}^{(l\sigma,\,2a)}\cdot\mathbf{\Omega},
\end{alignat}
and the friction tensors $\boldsymbol{\gamma}^{(l\sigma,\,1s)}$ and
$\boldsymbol{\gamma}^{(l\sigma,\,2a)}$ are given by
\begin{equation}
\boldsymbol{\gamma}^{(l\sigma,\,1s)}=\mathbf{P}^{(l\sigma)}\cdot\left[\mathbf{\boldsymbol{G}}^{-1}(\mathbf{R})\right]^{(l,\,1)}\cdot\mathbf{P}^{(1s)},\quad\boldsymbol{\gamma}^{(l\sigma,\,2a)}=\mathbf{P}^{(l\sigma)}\cdot\left[\mathbf{\boldsymbol{G}}^{-1}(\mathbf{R})\right]^{(l,\,2)}\cdot\mathbf{P}^{(2a)}.\label{eq:passive-friction-tensors}
\end{equation}
The friction tensors above give the contribution to the traction from
rigid body motion \cite{brenner1963,felderhof1976}. The traction
modes $l\sigma=1s$ and $l\sigma=2a$ correspond to the forces and
torques\begin{subequations}
\begin{eqnarray}
\mathbf{F}^{\mathcal{D}}=-\,\boldsymbol{\gamma}^{TT}\cdot\mathbf{V}-\boldsymbol{\gamma}^{TR}\cdot\mathbf{\Omega},\label{eq:traction-l-sigma-1-1}\\
\mathbf{T}^{\mathcal{D}}=-\,\boldsymbol{\gamma}^{RT}\cdot\mathbf{V}-\boldsymbol{\gamma}^{RR}\cdot\mathbf{\Omega}.
\end{eqnarray}
\end{subequations}where we have introduced the indices $T,R=1s,2a$
to make contact with the usual notation. The inverse of the single-layer
operator can be computed in several ways both analytically and numerically.
The Jacobi iteration used in \cite{singh2016crystallization,singh2016traction}
is convenient for analytical expressions. 

The fluid flow $\mathbf{u}^{\mathcal{D}}$ for a rigid body motion
of the sphere has the integral representation $\mathbf{\mathbf{u}^{\mathcal{D}}(\mathbf{r})}=-\,\boldsymbol{G}{}^{(l)}(\mathbf{R},\,\mathbf{r})\cdot\mathbf{F^{\mathcal{D}}}^{(l)}$
where the boundary integral
\begin{equation}
\boldsymbol{G}^{(l)}(\mathbf{r},\mathbf{R})=\frac{2l-1}{4\pi b^{2}}\int\mathbf{G}^{\text{}}(\mathbf{r},\,\mathbf{R}+\bm{\rho})\mathbf{Y}^{(l-1)}(\hat{\bm{\rho}})\,d\mathrm{S},\label{eq:rigid-body-flow}
\end{equation}
is the contribution to the external flow from the $l$-th tensorial
coefficient of the traction. The double layer has no contribution
from rigid body motion to the external flow. This result will be used
below in deriving a key identity necessary for deriving the variance
of the Brownian traction. 

\subsection{Active traction}

The linear algebraic system for the active contribution to the traction
is
\begin{alignat}{1}
 & -\boldsymbol{G}^{(l,\,l')}(\mathbf{R})\cdot\mathbf{F}^{\mathcal{A}(l')}+\boldsymbol{K}^{(l,\,l')}(\mathbf{R})\cdot\mathbf{V}^{(l')}=\tfrac{1}{2}\mathbf{V}^{(l)},\label{eq:linear-system-1-2}
\end{alignat}
where the matrix elements of the double-layer operator $\boldsymbol{K}(\mathbf{R)}$
in the tensorial harmonic basis are
\begin{equation}
\boldsymbol{K}^{(l,\,l')}(\mathbf{R})=\frac{2l-1}{4\pi b^{2}}\frac{1}{(l'-1)!(2l'-3)!!}\int\mathbf{Y}^{(l-1)}(\hat{\bm{\rho}})\mathbf{K}(\mathbf{R}+\bm{\rho},\,\mathbf{R}+\bm{\rho}')\cdot\hat{\bm{\rho}}'\,\mathbf{Y}^{(l'-1)}(\hat{\bm{\rho}}')\,d\mathrm{S}\,d\mathrm{S'}.
\end{equation}
The irreducible parts of the slip coefficients are $\mathbf{V}^{(l\sigma)}$.
The first two modes, $\mathbf{V}^{(1s)}\equiv-\mathbf{V}^{\mathcal{A}}$
and $\mathbf{V}^{(2a)}\equiv-b\mathbf{\Omega}^{\mathcal{A}}$ are
given by the integrals
\begin{alignat}{1}
4\pi a^{2}\,\mathbf{V}^{\mathcal{A}} & =-\int\boldsymbol{v}^{\mathcal{A}}(\bm{\rho})dS,\qquad4\pi a^{2}\,\bm{\Omega}^{\mathcal{A}}=-\frac{3}{2a^{2}}\int\bm{\rho}\times\boldsymbol{v}^{\mathcal{A}}(\bm{\rho})dS,
\end{alignat}
and are equal to the self-propulsion velocity and self-rotation angular
velocity of an isolated active colloid in unbounded flow \cite{anderson1989colloid,ghose2014irreducible}.
The solution of the linear system is
\begin{alignat}{1}
\mathbf{F}^{^{\mathcal{A}}(l\sigma)}=- & \sum_{l'\sigma'=1s}^{\infty}\,\boldsymbol{\gamma}^{(l\sigma,\,l'\sigma')}\cdot\mathbf{V}^{(l'\sigma')}.\label{eq:traction-l-sigma-2}
\end{alignat}
The generalized friction tensors $\boldsymbol{\gamma}^{(l\sigma,\,l'\sigma')}$
give the active contribution to the traction. These tensors were first
introduced in \cite{singh2016traction} and are given by
\begin{equation}
\boldsymbol{\gamma}^{(l\sigma,\,l'\sigma')}=\mathbf{P}^{(l\sigma)}\cdot\left[\boldsymbol{G}^{-1}\cdot\left(\tfrac{1}{2}\mathbf{I}-\boldsymbol{K}\right)\right]^{(l,\,l')}\cdot\mathbf{P}^{(l'\sigma')}.\label{eq:generalized-friction-tensor}
\end{equation}
They can be interpreted as Onsager coefficients of the linear response
of the traction to the surface slip. The above expression is identical
to Eq. (\ref{eq:passive-friction-tensors}) for $l'\sigma'=1s,2a$
since rigid body motion is an eigenvector of the double layer operator.
Therefore, both active \emph{and }passive friction tensors can be
recovered from the general expression above. 

The active force and torque on the colloid are the $l\sigma=1s$ and
$2a$ coefficients of the traction. These are given as\begin{subequations}
\begin{eqnarray}
\mathbf{F}^{\mathcal{A}}=-\sum_{l'\sigma'=1s}^{\infty}\,\boldsymbol{\gamma}^{(T,\,l'\sigma')}\cdot\mathbf{V}^{(l'\sigma')},\label{eq:traction-l-sigma-2-1}\\
\mathbf{T}^{\mathcal{A}}=-\sum_{l'\sigma'=1s}^{\infty}\,\boldsymbol{\gamma}^{(R,\,l'\sigma')}\cdot\mathbf{V}^{(l'\sigma')}.
\end{eqnarray}
\end{subequations}The active forces and torques depend on all modes
of the slip. In contrast to passive colloids, where only four friction
tensors determine the force and torque, there are, in general, infinitely
many friction tensors for active colloids, corresponding to the infinitely
many modes of the slip. These infinitely many friction tensors account
for the diversity of phenomenon seen in active suspensions.

\subsection{Brownian traction}

The linear algebraic system for the Brownian contribution to the traction
is
\begin{alignat}{1}
\boldsymbol{G}{}^{(l,\,l')}(\mathbf{R})\cdot\hat{\mathbf{F}}^{(l')}(t) & =\mathbf{\boldsymbol{W}}^{(l)}(\mathbf{R},\,t),\label{eq:fluctuationEq}
\end{alignat}
where $\mathbf{\boldsymbol{W}}^{(l)}$ are coefficients of the irreducible
tensor expansion
\begin{equation}
\boldsymbol{w}(\mathbf{r)}=\sum_{l=1}^{\infty}\frac{1}{(l-1)!(2l-3)!!}\,\mathbf{\boldsymbol{W}}^{(l)}\cdot\mathbf{Y}^{(l-1)}(\bm{\hat{\rho}}),\qquad\text{on }S,
\end{equation}
of the Brownian velocity field incident on the surface of the colloid.
From the definition of the Brownian velocity field, Eq.(\ref{eq:Brownian-velocity-field}),
these coefficients are given by
\begin{equation}
\mathbf{\boldsymbol{W}}^{(l)}=\int\boldsymbol{G}{}^{(l)}\cdot\boldsymbol{\mathbf{\xi}}(\mathbf{r}')\,dV'.
\end{equation}
The solution of the linear system for the Brownian traction is
\begin{equation}
\hat{\mathbf{F}}^{(l)}(t)=\left[\mathbf{\boldsymbol{G}}^{-1}(\mathbf{R})\right]^{(l,\,l')}\cdot\mathbf{\boldsymbol{W}}^{(l')}(\mathbf{R},\,t).
\end{equation}
The coefficients of the Brownian traction are proportional to the
coefficients of the Brownian velocity field incident on the surface
of the colloid and, by Eq. (\ref{eq:Brownian-velocity-field}), to
the thermal force in the fluctuating Stokes equation. It is clear,
then, that the traction coefficients are zero-mean Gaussian random
variables and to fully specify their distribution it is necessary,
then, to only determine their variance. By the previous equation,
their variance is related to that of the Brownian velocity coefficients
as
\begin{alignat}{1}
\langle\hat{\mathbf{F}}^{(l)}(t) & \hat{\mathbf{F}}^{(l')}(t')\rangle=\left[\mathbf{\boldsymbol{G}}^{-1}(\mathbf{R})\right]^{(l,\,k)}\,\big\langle\mathbf{\boldsymbol{W}}^{(k)}(\mathbf{R},\,t')\mathbf{\boldsymbol{W}}^{(k')}(\mathbf{R},\,t')\rangle\,\left[\mathbf{\boldsymbol{G}}^{-1}(\mathbf{R})\right]^{(l',\,k')}.\label{eq:var-Fhat}
\end{alignat}

To determine the variance of Brownian velocity coefficients we use
the boundary integral representation of $\boldsymbol{u}^{\mathcal{D}}$
given above (see also \cite{singh2015many,singh2016traction}). Inserting
this on the left of the fluctuation-dissipation relation for the thermal
force, Eq. (\ref{eq:FDT-Hauge}), gives
\begin{alignat}{1}
\Big\langle\int\mathbf{u}^{\mathcal{D}}(\mathbf{r)}\cdot\boldsymbol{\xi}(\mathbf{r},\,t)\,dV\int\mathbf{u}^{\mathcal{D}}(\mathbf{r')}\cdot\boldsymbol{\xi}(\mathbf{r}',\,t')\,dV'\Big\rangle=\mathbf{F}^{\mathcal{D}(l)}\cdot\Big\langle\mathbf{\boldsymbol{W}}^{(l)}(\mathbf{R},\,t')\mathbf{\boldsymbol{W}}^{(l')}(\mathbf{R},\,t')\Big\rangle\cdot\mathbf{F}^{\mathcal{D}(l')}.
\end{alignat}
On the other hand, the power dissipated by the rigid body motion,
on the right of the fluctuation-dissipation relation can be expressed
as
\begin{equation}
\dot{\mathcal{E}}(\mathbf{u}^{\mathcal{D}})=-\int\boldsymbol{f}^{\mathcal{D}}(\mathbf{R}+\mathbf{\boldsymbol{\rho}})\cdot\mathbf{u}^{\mathcal{D}}(\mathbf{R}+\mathbf{\boldsymbol{\rho}})d\text{S}=-\mathbf{F}^{\mathcal{D}(l)}\cdot\mathbf{V}^{\mathcal{D}(l)}=\mathbf{F}^{\mathcal{D}(l)}\cdot\boldsymbol{G}{}^{(l,\,l')}(\mathbf{R})\cdot\mathbf{F}^{\mathcal{D}(l')}.\label{eq:power-disspation}
\end{equation}
The first equality is obtained by using the divergence theorem to
reduce the volume integral of the quadratic form in Eq. (\ref{eq:dissipation-W})
to the boundary of the colloid and then using the constitutive relation
between the stress and the strain rate in Stokes flow \cite{landau1959fluid}.
The second equality is an elementary consequence of the orthogonality
of the tensorial spherical harmonics \cite{singh2015many,singh2016traction}.
The third equality is obtained by eliminating the velocity using the
linear algebraic system, Eq.(\ref{eq:linear-system-passive}) for
rigid body motion. Comparing the above two equations, we obtain the
key identity for the variance of the Brownian velocity coefficients,
\begin{equation}
\langle\mathbf{\boldsymbol{W}}^{(l)}(\mathbf{R},\,t')\mathbf{\boldsymbol{W}}^{(l')}(\mathbf{R},\,t')\rangle=2k_{B}T\,\boldsymbol{G}^{(l,\,l')}(\boldsymbol{R})\,\delta(t-t').
\end{equation}
This expression when used in Eq. (\ref{eq:var-Fhat}) yields the variance
of Brownian traction coefficients
\begin{alignat}{1}
\langle\hat{\mathbf{F}}^{(l)}(t) & \hat{\mathbf{F}}^{(l')}(t')\rangle=2k_{B}T\,\big[\mathbf{\boldsymbol{G}}^{-1}(\mathbf{R})\big]^{(l,\,l')}\delta(t-t').\label{eq:avg-lth-brow}
\end{alignat}
These are an infinite number of fluctuation-dissipation relations
between the variance of the tensorial harmonic modes of the fluctuating
traction and the matrix elements, in the irreducible tensorial harmonic
basis, of the \emph{inverse }of the single-layer operator. To the
best of our knowledge, the complete statistics of the Brownian traction
is derived here for the first time and is the central result of this
paper. 

The variance of the irreducible parts of the fluctuating traction
follow straightforwardly as
\begin{alignat}{1}
\langle\hat{\mathbf{F}}^{(l\sigma)}(t) & \hat{\mathbf{F}}^{(l'\sigma')}(t')\rangle=2k_{B}T\,\mathbf{P}^{(l\sigma)}\cdot\big[\mathbf{\boldsymbol{G}}^{-1}(\mathbf{R})\big]^{(l,\,l')}\cdot\mathbf{P}^{(l'\sigma')}\,\delta(t-t').\label{eq:general-fdt}
\end{alignat}
The first two coefficients of the fluctuating traction are the force
and torque, and choosing $l\sigma=1s,\,2a$ we obtain\begin{subequations}
\begin{align}
\langle\hat{\mathbf{F}}\rangle & =0,\qquad\langle\hat{\mathbf{F}}(t)\,\hat{\mathbf{F}}(t')\rangle=2k_{B}T\,\boldsymbol{\gamma}^{TT}\delta(t-t'),\qquad\langle\hat{\mathbf{F}}(t)\,\hat{\mathbf{T}}(t')\rangle=2k_{B}T\,\boldsymbol{\gamma}^{TR}\delta(t-t'),\\
\langle\hat{\mathbf{T}}\rangle & =0,\qquad\langle\hat{\mathbf{T}}(t)\,\hat{\mathbf{F}}(t')\rangle=2k_{B}T\,\boldsymbol{\gamma}^{RT}\delta(t-t'),\qquad\langle\hat{\mathbf{T}}(t)\,\hat{\mathbf{T}}(t')\rangle=2k_{B}T\,\boldsymbol{\gamma}^{RR}\delta(t-t').
\end{align}
\end{subequations}where $\boldsymbol{\gamma}^{\alpha\beta}$, with
$\alpha,\beta=T,R$, are the one-particle friction tensor and are
($l\sigma=1s,\,2a$) elements of $\mathbf{P}^{(l\sigma)}\cdot\big[\mathbf{\boldsymbol{G}}^{-1}(\mathbf{R})\big]^{(l,\,l')}\cdot\mathbf{P}^{(l'\sigma')}$.
The fluctuation-dissipation relation for the Brownian force and torque
are, thus, derived from the fluctuation-dissipation relation for the
thermal force on the fluid.

We make the following remarks about the above derivation. First, the
explicit form of the inverse of the single-layer operator is is not
necessary to obtain Eq. (\ref{eq:avg-lth-brow}); it is sufficient
to know that the inverse exists. Therefore, the fluctuation-dissipation
relation for the irreducible coefficients, Eq. (\ref{eq:general-fdt}),
is valid for any geometry bounding the fluid, provided the flow vanishes
there. In particular, it holds for a colloid near a plane wall. Second,
our derivation provides the fluctuation-dissipation relation for \emph{all
}modes of the Brownian traction. Earlier derivations have focused
on only the force and torque. Therefore, our derivation provides the
fluctuating stresslet ($l\sigma=2s)$ which is needed to compute the
Brownian contribution to the suspension stress. Third, the configuration-dependence
of both the fluctuation, Eq. (\ref{eq:fluctuationEq}), and the dissipation,
Eq. (\ref{eq:general-fdt}), is made explicit in our derivation. The
configuration-dependent ``noise'' variance follows from this automatically.
The interpretation of the resulting multiplicative noise in the Langevin
equation that we derive below is obtained by recalling that the momentum
and angular momentum of the colloid are fast variables that have,
implicitly, been adiabatically eliminated \cite{gardiner1984adiabatic}.
The form of the Smoluchowski equation for this problem is well-known
and is used below to consistently interpret the multiplicative noise
in the Langevin equation \cite{chandrasekhar1949brownian,murphy1972brownian,wilemski1976derivation}.

\section{Langevin equations for a spherical active colloid \label{sec:Langevin-equations- }}

In this section we derive the Langevin and Smoluchowski equations
for the Brownian motion of an active colloid. We use the results derived
above for three kinds of forces and torques acting on the colloid.
With these, the balance conditions, Eq. (\ref{eq:newtons}), become\begin{subequations}\label{force-formulation-1-1}
\begin{alignat}{1}
-\boldsymbol{\gamma}^{TT}\mathbf{\cdot V}-\boldsymbol{\gamma}^{TR}\mathbf{\cdot\,\boldsymbol{\Omega}}-\sum_{l\sigma=1s}^{\infty}\boldsymbol{\gamma}^{(T,\,l\sigma)}\cdot\mathbf{V}^{(l\sigma)}+\hat{\mathbf{F}}+\mathbf{F}^{P}=0,\\
-\boldsymbol{\gamma}^{RT}\mathbf{\cdot V}-\boldsymbol{\gamma}^{RR}\mathbf{\cdot\,\boldsymbol{\Omega}}-\sum_{l\sigma=1s}^{\infty}\boldsymbol{\gamma}^{(R,\,l\sigma)}\cdot\mathbf{V}^{(l\sigma)}+\hat{\mathbf{T}}+\mathbf{T}^{P}=0.
\end{alignat}
\end{subequations}The above can be inverted to obtain the rigid body
motion of the colloid in explicit form. This gives the Langevin equations
for a Brownian active colloid with hydrodynamic interactions, first
derived heuristically in \cite{laskar2015brownian},\begin{subequations}\label{eq:mobility-formulation}
\begin{alignat}{1}
\mathsf{\mathbf{V}} & =\boldsymbol{\mu}^{TT}\cdot\mathbf{F}^{P}+\boldsymbol{\mu}^{TR}\cdot\mathbf{T}^{P}+\sqrt{2k_{B}T\bm{\mu}^{TT}}\cdot\bm{\eta}^{T}+\sqrt{2k_{B}T\bm{\mu}^{TR}}\cdot\bm{\zeta}^{R}+\sum_{l\sigma=2s}^{\infty}\boldsymbol{\pi}^{(T,\,l\sigma)}\cdot\mathbf{\mathsf{\mathbf{V}}}^{(l\sigma)}+\mathsf{\mathbf{V}}^{\mathcal{A}},\label{eq:RBM-velocity}\\
\mathsf{\mathbf{\Omega}} & =\underbrace{\boldsymbol{\mu}^{RT}\cdot\mathbf{F}^{P}+\boldsymbol{\mu}^{RR}\cdot\mathbf{T}^{P}}_{\mathrm{Passive}}+\underbrace{\sqrt{2k_{B}T\bm{\mu}^{RT}}\cdot\bm{\zeta}^{T}+\sqrt{2k_{B}T\bm{\mu}^{RR}}\cdot\bm{\eta}^{R}}_{\mathrm{Brownian}}+\sum_{l\sigma=2s}^{\infty}\underbrace{\boldsymbol{\pi}^{(R,\,l\sigma)}\cdot\mathbf{\mathsf{\mathbf{V}}}^{(l\sigma)}+\mathsf{\mathbf{\Omega}}^{\mathcal{A}}}_{\mathrm{Active}}.\label{eq:RBM-angular-velocity}
\end{alignat}
\end{subequations}Here $\boldsymbol{\eta}^{\alpha}$, and $\boldsymbol{\zeta}^{\alpha}$
are Gaussian white noises with zero-mean and variances $1$ and $1/b$
respectively. The matrix square roots are to be interpreted as Cholesky
factors. The mobility matrices $\boldsymbol{\mu}^{\alpha\beta}$ are
inverses of the friction matrices $\boldsymbol{\gamma}^{\alpha\beta}$
\cite{happel1965low,felderhof1977hydrodynamic,mazur1982,schmitz1982mobility,nunan1984effective,ladd1988,durlofsky1987dynamic,brady1988dynamic,kim2005,cichocki1994friction}.
The propulsion tensors $\boldsymbol{\pi}^{(\alpha,\,l\sigma)}$, first
introduced in \cite{singh2015many}, relate the rigid body motion
to modes of the active velocity. They are related to the generalized
friction tensors, introduced in \cite{singh2016traction}, by\begin{subequations}\label{eq:pi-mu-relation-1}
\begin{alignat}{1}
-\bm{\pi}^{(\text{T},\,l\sigma)} & =\boldsymbol{\mu}^{TT}\cdot\boldsymbol{\gamma}^{(T,\,l\sigma)}+\boldsymbol{\mu}^{TR}\cdot\boldsymbol{\gamma}^{(R,\,l\sigma)},\\
-\bm{\pi}^{(R,\,l\sigma)} & =\boldsymbol{\mu}^{RT}\cdot\boldsymbol{\gamma}^{(T,\,l\sigma)}+\boldsymbol{\mu}^{RR}\cdot\boldsymbol{\gamma}^{(R,\,l\sigma)}.
\end{alignat}
\end{subequations} The translational propulsion tensors $\bm{\pi}^{(\text{T},\,l\sigma)}$
are dimensionless while the rotational propulsion tensors $\bm{\pi}^{(R,\,l\sigma)}$
have dimensions of inverse length. Stochastic trajectories of motion
can be obtained from the kinematic equations\textcolor{black}{
\begin{equation}
\dot{\mathbf{R}}=\mathbf{V},\quad\dot{\mathbf{p}}=\boldsymbol{\Omega}\times\mathbf{p},
\end{equation}
}using the standard Ermak-McCammon integrator \cite{ermak1978}. In
this integrator, the noise variances are computed using mobilities
in the configuration at time $t$ but a ``spurious'' drift, proportional
to the configurational divergence of the mobilities, is added to to
arrive at the configuration at time $t+\Delta t$. There is nothing
particularly spurious about this drift; it is simply the residual
effect of the adiabatically eliminated degrees of freedom. 

The Smoluchowski equation for the distribution function $\Psi(\mathbf{R};\,\mathbf{p})$
of positions and orientations follows immediately from the Langevin
equations. We write it in the form of a conservation law in configuration
space
\begin{align}
\frac{\partial\Psi}{\partial t} & =\boldsymbol{\mathcal{L}}\Psi=-\left(\boldsymbol{\nabla}_{{\scriptscriptstyle \mathbf{R}}}\cdot\boldsymbol{\mathcal{V}}_{{\scriptscriptstyle \mathbf{R}}}+\mathbf{p}\times\boldsymbol{\nabla}_{{\scriptscriptstyle \mathbf{p}}}\cdot\boldsymbol{\mathcal{V}}_{{\scriptscriptstyle \mathbf{p}}}\right)\Psi,\label{eq:dist-func}
\end{align}
where the ``velocities'' $\boldsymbol{\mathcal{V}}_{{\scriptscriptstyle \mathbf{R}}}$
and $\boldsymbol{\mathcal{V}}_{{\scriptscriptstyle \mathbf{p}}}$
are\begin{subequations}
\begin{align}
\boldsymbol{\mathcal{V}}_{{\scriptscriptstyle \mathbf{R}}} & =\boldsymbol{\mu}^{TT}\cdot\left(\mathbf{F}^{P}-k_{B}T\,\boldsymbol{\nabla}_{{\scriptscriptstyle \mathbf{R}}}\right)+\boldsymbol{\mu}^{TR}\cdot\left(\mathbf{T}^{P}-k_{B}T\,\mathbf{p}\times\boldsymbol{\nabla}_{{\scriptscriptstyle \mathbf{p}}}\right)+\sum_{l\sigma=2s}^{\infty}\boldsymbol{\pi}^{(T,\,l\sigma)}\cdot\mathbf{V}^{(l\sigma)}+\mathbf{V}^{\mathcal{A}},\\
\boldsymbol{\mathcal{V}}_{{\scriptscriptstyle \mathbf{p}}} & =\boldsymbol{\mu}^{RT}\cdot\left(\mathbf{F}^{P}-k_{B}T\,\boldsymbol{\nabla}_{{\scriptscriptstyle \mathbf{R}}}\right)+\boldsymbol{\mu}^{RR}\cdot\left(\mathbf{T}^{P}-k_{B}T\,\mathbf{p}\times\boldsymbol{\nabla}_{{\scriptscriptstyle \mathbf{p}}}\right)+\sum_{l\sigma=2s}^{\infty}\boldsymbol{\pi}^{(R,\,l\sigma)}\cdot\mathbf{V}^{(l\sigma)}+\mathbf{\Omega}^{\mathcal{A}}.
\end{align}
\end{subequations}Here $\mathbf{F}^{P}=-\boldsymbol{\nabla}_{{\scriptscriptstyle \mathbf{R}}}U$,
$\mathbf{T}^{P}=-\mathbf{p}\times\boldsymbol{\nabla}_{{\scriptscriptstyle \mathbf{p}}}U$,
and $U$ is a potential that contains both positional and orientational
interactions. Note the position of the mobility in the second derivative
terms: the $\boldsymbol{\nabla\mu\nabla}$ order (in contrast with
two other inequivalent permutations) is provided unambiguously by
the adiabatic elimination of momenta. 

In the absence of activity, the drift and diffusion coefficients in
the Smoluchowski equation obey the fluctuation-dissipation relation
and the Gibbs distribution $\Psi\sim\exp(-U/k_{B}T$) is the stationary
solution. However, the Gibbsian form is not a stationary solution
in the presence of the active terms, as can easily be verified by
substitution. This applies, a fortiori, to an active colloid near
a plane wall discussed in the next section, where the barometric distribution
of a passive suspension is no longer the stationary distribution.
\begin{figure}
\includegraphics[width=0.98\textwidth]{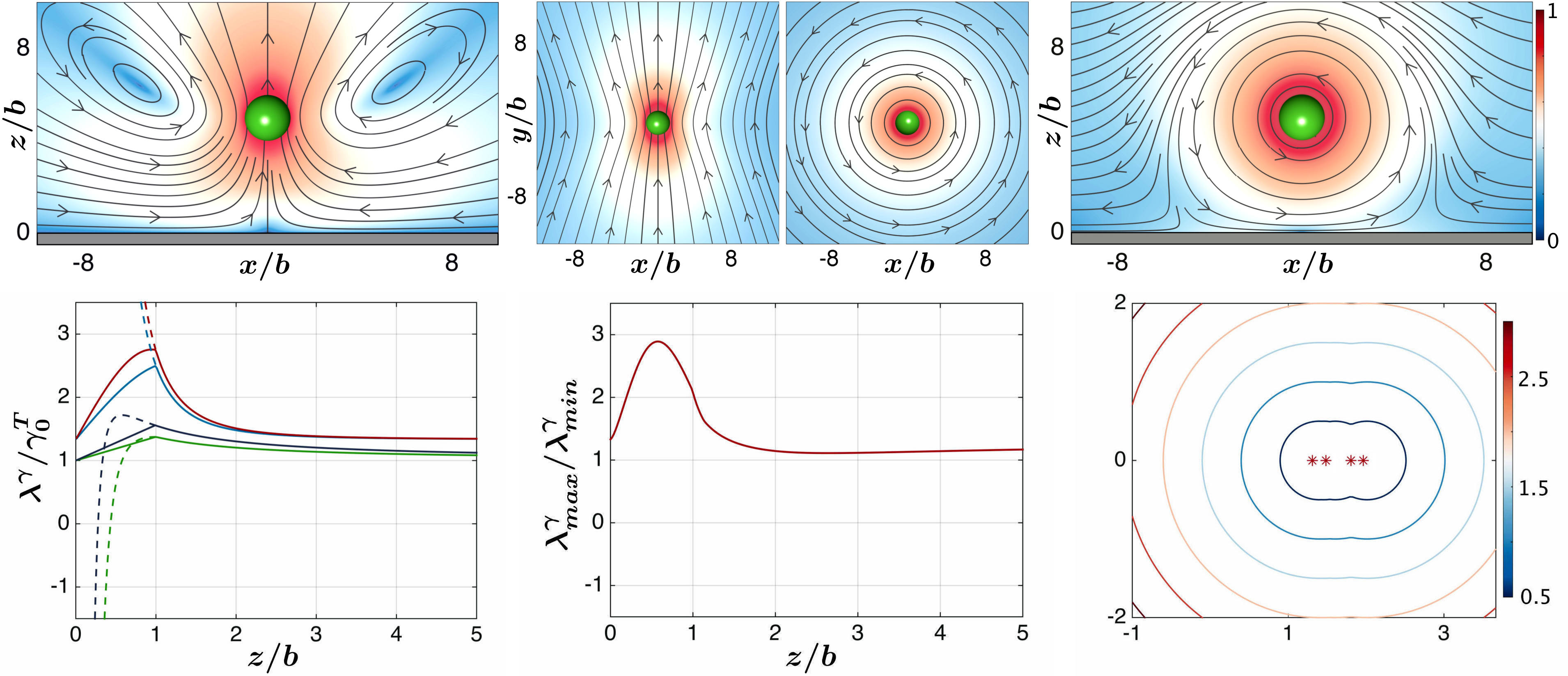}

\caption{Fluid flow due to rigid body motion and spectral properties of friction
tensors for a spherical colloid near a plane wall. The first two figures
in the top panel show the flow due to translation perpendicular and
parallel to the wall. The next two figures show the flow due to rotation
perpendicular and parallel to the wall. The first figure in the bottom
panel shows eigenvalues, normalized by $\gamma_{0}^{T}$, of the $6\times6$
grand friction tensor assembled out of the $3\times3$ blocks in Table
(\ref{tab:gamma-ab}). The dashed lines show the eigenvalues of the
unregularised tensor and its loss of positive-definiteness for $h<b$.
The next two figures show the condition number of the grand friction
tensor and its pseudo-spectrum in the complex plane when $h=1.3b$.
Here $\boldsymbol{\gamma}_{0}^{T}=6\pi\eta b$ and $\boldsymbol{\gamma}_{0}^{R}=8\pi\eta b^{3}$
are, respectively, the friction for translation and rotation in an
unbounded fluid.\label{fig:1}}
\end{figure}

\section{Brownian active colloid near a plane wall\label{sec:wall-Bounded}}

We now apply the preceding general results to the specific case of
an active colloid near a plane wall. The Green's function for the
problem, denoted by $\mathbf{G}^{\text{w}}$, is taken to vanish at
the location of the wall, $z=0$. The form of the Green's function,
first derived by Lorentz \cite{lorentz1896eene}, can be written in
the following form due to Blake \cite{blake1971c}
\begin{alignat}{1}
G_{\alpha\beta}^{\text{w}}(\mathbf{R}',\,\mathbf{R}) & =G_{\alpha\beta}^{0}(\mathbf{R}'-\mathbf{R})+G_{\alpha\beta}^{*}(\mathbf{R}',\,\mathbf{R}^{*}),\label{eq:wall-G}
\end{alignat}
where $G_{\alpha\beta}^{0}(\mathbf{r})=(\nabla^{2}\delta_{\alpha\beta}-\nabla_{\alpha}\nabla_{\beta})(r/8\pi\eta$)
is the Oseen tensor and the correction necessary to satisfy the boundary
condition is\textcolor{black}{
\begin{alignat}{1}
G_{\alpha\beta}^{*} & =\frac{1}{8\pi\eta}\left[-\frac{\delta_{\alpha\beta}}{r^{*}}-\frac{r_{\alpha}^{*}r_{\beta}^{*}}{r^{*^{3}}}+2h^{2}\bigg(\frac{\delta_{\alpha\nu}}{r^{*^{3}}}-\frac{3r_{\alpha}r_{\nu}}{r^{*^{5}}}\bigg)\mathcal{M}_{\nu\beta}-2h\bigg(\frac{r_{3}^{*}\delta_{\alpha\nu}+\delta_{\nu3}r_{\alpha}^{*}-\delta_{\alpha3}r_{\nu}^{*}}{r^{*3}}-\frac{3r_{\alpha}r_{\nu}r_{3}^{*}}{r^{*^{5}}}\bigg)\mathcal{M}_{\nu\beta}\right].
\end{alignat}
}The correction is interpreted as a sum of three images, a Stokeslet,
a dipole, and a degenerate quadrupole, located at $\mathbf{\mathbf{R}}^{*}=\boldsymbol{\mathcal{M}}\cdot\mathbf{\mathbf{R}}$,
where $\boldsymbol{\mathcal{M}}=\boldsymbol{I}-2\mathbf{\hat{z}}\mathbf{\hat{z}}$
is the mirror operator with respect to the wall \textcolor{black}{\cite{brennen1977,berke2008hydrodynamic,lauga2009,goldstein2015green,mathijssen2015tracer,singh2016crystallization}}.
Here $h$ is the height of the colloid from the wall and $\mathbf{r}^{*}=\mathbf{\mathbf{R}}'-\mathbf{\mathbf{R}}^{*}$.
The correction has no singularities in the domain of flow and the
only singular contribution there is from the Oseen tensor. The flows
produced by a sphere translating or rotating near the wall are shown
in the top panels of Fig. (\ref{fig:1}). 

The leading terms of the inverse of the single-layer operator are
obtained here using Jacobi's iterative method \cite{singh2016traction}.
The results for the friction tensors $\boldsymbol{\gamma}^{\alpha\beta}$
are given in the left column of Table (\ref{tab:gamma-ab}). The grand
resistance tensor, formed out each of the $3\times3$ blocks, is manifestly
symmetric and, for $h>b$, positive-definite. This is established
by an explicit computation of the eigenvalues, shown in the first
figure of the bottom panel of Fig. (\ref{fig:1}). Thus positive-definiteness
is ensured for all configurations in which the colloid does not overlap
with the wall. 

In Brownian dynamics simulations, however, it is convenient to avoid
hard-sphere potentials as these require special integrators. In the
absence of such potentials, it is no longer possible to maintain the
constraint $h>b$ during integration, and the colloid may substantially
overlap with the wall. For such configurations, a naive continuation
of the friction tensors to the domain $h<b$ is untenable: as the
dashed lines in the eigenvalue plot in Fig. (\ref{fig:1}) show, the
grand resistance tensor has negative eigenvalues in this region. This
implies negative entropy production and is clearly unphysical. The
cure, first proposed by Rotne and Prager in the context of bead-spring
models of polymers \cite{rotne1969}, is to regularize the matrix
elements of the single-layer, by computing the surface integrals in
their definition, over the union of overlapping surfaces. In this
case, the integral is to be computed over the union of the surface
of the colloid and its image. Such overlap integrals have been computed
recently \cite{wajnryb2013generalization} and we use those results
to obtain the regularized forms of the friction in the right column
of Table (\ref{tab:gamma-ab}). The use of the regularized form provides
friction tensors that are positive-definite for $all$ heights, as
can be seen from the solid curves for $h<b$ in the eigenvalue plot
in Fig. (\ref{fig:1}). Their use results in positive-definite variances
for the Brownian force and torque at $all$ heights above the wall. 

In the second and third figures of the bottom panel, we plot the condition
number of the grand resistance tensor as a function of height and
its pseudo-spectrum at $h=1.3b$. The condition number remains small
for all heights and the pseudo-spectrum shows no sign of non-normality.
Thus, computing the ``square-root'' Cholesky factors of the grand
friction tensor (or, its inverse, the grand mobility tensor) poses
no problem and iterative methods are expected to converge rapidly.
Brownian dynamics simulations of active colloids near a plane wall
can then be efficiently performed using the above results.

\begin{table}[t]
\centering\renewcommand{\arraystretch}{2} 

\begin{tabular}{|>{\centering}m{8cm}|>{\centering}m{9cm}|}
\hline 
$h>b$ (Jacobi iteration) & $h\leq b$ (Regularized)\tabularnewline
\hline 
\hline 
$\boldsymbol{\gamma}^{TT}=\gamma_{0}^{T}\big(\boldsymbol{I}-\gamma_{0}^{T}\mathcal{F}^{1}\mathcal{F}^{1}\mathbf{G}^{*}\big)$ & $\boldsymbol{\gamma}^{TT}=\gamma_{0}^{T}\left[\big(1+\tfrac{9r^{*}}{32b}\big)\boldsymbol{I}-\frac{3\mathbf{r}^{*}\mathbf{r}^{*}}{32br^{*}}\right]$\tabularnewline
\hline 
$\boldsymbol{\gamma}^{RR}=\gamma_{0}^{R}\big(\boldsymbol{I}-\tfrac{\gamma_{0}^{R}}{4}\boldsymbol{\nabla}_{{\scriptscriptstyle \mathbf{R}}}\times\boldsymbol{\nabla}_{{\scriptscriptstyle \mathbf{R}}}\times\mathbf{G}^{*}\big)$ & $\boldsymbol{\gamma}^{RR}=\gamma_{0}^{R}\Big[\big(1+\tfrac{27r^{*}}{32b}-\tfrac{27r^{*^{3}}}{64b^{3}}\big)\boldsymbol{I}-\big(\tfrac{9r}{32b}-\tfrac{3r^{*^{3}}}{64b^{3}}\big)\frac{\mathbf{r}^{*}\mathbf{r}^{*}}{r^{*2}}\Big]$\tabularnewline
\hline 
$\boldsymbol{\gamma}^{RT}=-\tfrac{\gamma_{0}^{R}\gamma_{0}^{T}}{2}\boldsymbol{\nabla}_{{\scriptscriptstyle \mathbf{R}}}\times\mathbf{G}^{*}$ & $\boldsymbol{\gamma}^{RT}=-\gamma_{0}^{R}\big(\tfrac{2}{b^{2}}-\tfrac{3r^{*}}{4b^{3}}\big)\boldsymbol{\varepsilon}\cdot\mathbf{r}^{*}$\tabularnewline
\hline 
\end{tabular}\caption{\textcolor{black}{Expressions for the $3\times3$ friction tensors
at a height $h$ from a plane wall,} using the result of the first
Jacobi iteration \cite{singh2016traction} for $h>b$ and its regularized
form for $h\leq b$.\textcolor{black}{{} Here }$\mathcal{F}^{1}=1+\tfrac{b^{2}}{6}\nabla_{{\scriptscriptstyle \mathbf{R}}}^{2}$
is an operator encoding the finite size of the colloid.\label{tab:gamma-ab} }
\end{table}

\section{Discussion}

We have shown how to obtain the three parts of the traction, due to
rigid body motion, activity and thermal fluctuations, on a spherical
colloid in a fluctuating Stokesian fluid. We then applied the general
results to the specific case of an active colloid near a plane wall.
We provided a regularization of the friction tensors to enable Brownian
dynamics simulations in which the spheres can overlap with the walls.
We conclude with three remarks about our results. 

The first is that expressions in the left column of Table (\ref{tab:gamma-ab})
are valid not only for the Lorentz-Blake Green's function, but for
any Green's function that can be expressed as $\mathbf{G}^{0}+\mathbf{G}^{\ast}$,
where $\mathbf{G}^{0}$ is the Oseen tensor and $\mathbf{G}^{*}$
is the correction necessary to satisfy the boundary conditions. A
variety of Green's function can be expressed in this form, including
those for flow between parallel walls and in periodic domains. As
mentioned before, the correction term does not contain singularities
in the domain of flow and, therefore, the boundary integrals in the
definition of the matrix elements can be expressed in terms of derivatives
of the correction. Singular terms from the Oseen tensor can be calculated
explicitly using well-known results for integrals of Bessel functions.
Therefore, our results for $h>b$ are of broader validity than might
have been anticipated. 

The second is that active colloids show fascinating behaviour in the
proximity of a wall \cite{palacci2013living,petroff2015fast}. While
Brownian motion appears to be negligible in comparison to activity-induced
motion for a large class of these colloids, there is considerable
theoretical interest in understanding the interplay between passive
friction, thermal fluctuations, and activity, especially when the
friction is configuration-dependent. Brownian dynamics, with the regularized
frictions provided here, will be a powerful tool to study such interplays. 

Finally, the extension of the method presented here to determine the
tractions on the surface of many active colloids is straightforward
in principle but tedious in practice. The result for the active traction
has been obtained recently \cite{singh2016traction} and a heuristic
argument has been provided to determine the first two moments of the
Brownian traction. It will be instructive to obtain all moments and,
in particular, the symmetric second moment to determine their contribution
to the stress in a non-dilute active Brownian suspension. 

\end{document}